\documentclass[twoside,twocolumn,9pt]{article}
\usepackage{extsizes}
\usepackage[super,sort&compress,comma]{natbib} 
\usepackage[version=3]{mhchem}
\usepackage[left=1.5cm, right=1.5cm, top=1.785cm, bottom=2.0cm]{geometry}
\usepackage{balance}
\usepackage{mathptmx}
\usepackage{sectsty}
\usepackage{graphicx} 
\usepackage{lastpage}
\usepackage[format=plain,justification=justified,singlelinecheck=false,font={stretch=1.125,small,sf},labelfont=bf,labelsep=space]{caption}
\usepackage{float}
\usepackage{fancyhdr}
\usepackage{fnpos}
\usepackage[english]{babel}
\addto{\captionsenglish}{%
  
}
\usepackage{array}
\usepackage{droidsans}
\usepackage{charter}
\usepackage[T1]{fontenc}
\usepackage[usenames,dvipsnames]{xcolor}
\usepackage{setspace}
\usepackage[compact]{titlesec}
\usepackage{hyperref} 
\usepackage{siunitx} 
\DeclareSIUnit{\au}{a.u.}

\usepackage{epstopdf}

\definecolor{cream}{RGB}{222,217,201}

\begin{document}

\pagestyle{fancy}
\thispagestyle{plain}
\fancypagestyle{plain}{
\renewcommand{\headrulewidth}{0pt}
}

\makeFNbottom
\makeatletter
\renewcommand\LARGE{\@setfontsize\LARGE{15pt}{17}}
\renewcommand\Large{\@setfontsize\Large{12pt}{14}}
\renewcommand\large{\@setfontsize\large{10pt}{12}}
\renewcommand\footnotesize{\@setfontsize\footnotesize{7pt}{10}}
\makeatother

\renewcommand{\thefootnote}{\fnsymbol{footnote}}
\renewcommand\footnoterule{\vspace*{1pt}%
\color{cream}\hrule width 3.5in height 0.4pt \color{black}\vspace*{5pt}} 
\setcounter{secnumdepth}{5}

\makeatletter 
\renewcommand\@biblabel[1]{#1}            
\renewcommand\@makefntext[1]%
{\noindent\makebox[0pt][r]{\@thefnmark\,}#1}
\makeatother 
\renewcommand{\figurename}{\small{Fig.}~}
\sectionfont{\sffamily\Large}
\subsectionfont{\normalsize}
\subsubsectionfont{\bf}
\setstretch{1.125} 
\setlength{\skip\footins}{0.8cm}
\setlength{\footnotesep}{0.25cm}
\setlength{\jot}{10pt}
\titlespacing*{\section}{0pt}{4pt}{4pt}
\titlespacing*{\subsection}{0pt}{15pt}{1pt}

\fancyfoot{}
\fancyfoot[LO,RE]{\vspace{-7.1pt}\includegraphics[height=9pt]{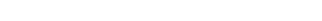}}
\fancyfoot[CO]{\vspace{-7.1pt}\hspace{11.9cm}\includegraphics{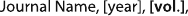}}
\fancyfoot[CE]{\vspace{-7.2pt}\hspace{-13.2cm}\includegraphics{head_foot/RF}}
\fancyfoot[RO]{\footnotesize{\sffamily{1--\pageref{LastPage} ~\textbar  \hspace{2pt}\thepage}}}
\fancyfoot[LE]{\footnotesize{\sffamily{\thepage~\textbar\hspace{4.65cm} 1--\pageref{LastPage}}}}
\fancyhead{}
\renewcommand{\headrulewidth}{0pt} 
\renewcommand{\footrulewidth}{0pt}
\setlength{\arrayrulewidth}{1pt}
\setlength{\columnsep}{6.5mm}
\setlength\bibsep{1pt}

\makeatletter 
\newlength{\figrulesep} 
\setlength{\figrulesep}{0.5\textfloatsep} 

\newcommand{\topfigrule}{\vspace*{-1pt}%
\noindent{\color{cream}\rule[-\figrulesep]{\columnwidth}{1.5pt}} }

\newcommand{\botfigrule}{\vspace*{-2pt}%
\noindent{\color{cream}\rule[\figrulesep]{\columnwidth}{1.5pt}} }

\newcommand{\dblfigrule}{\vspace*{-1pt}%
\noindent{\color{cream}\rule[-\figrulesep]{\textwidth}{1.5pt}} }

\makeatother

\twocolumn[
  \begin{@twocolumnfalse}
{\includegraphics[height=30pt]{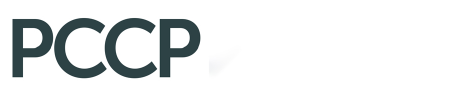}\hfill\raisebox{0pt}[0pt][0pt]{\includegraphics[height=55pt]{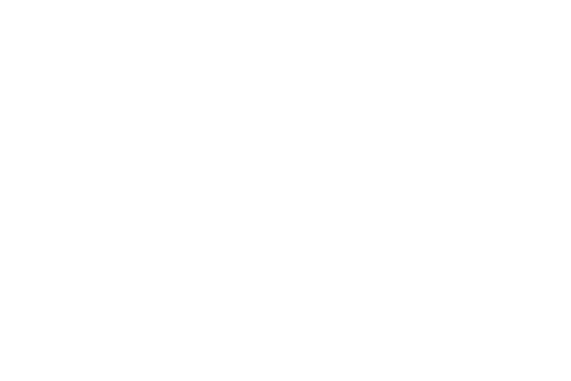}}\\[1ex]
\includegraphics[width=18.5cm]{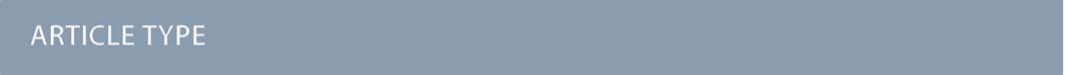}}\par
\vspace{1em}
\sffamily
\begin{tabular}{m{4.5cm} p{13.5cm} }

\includegraphics{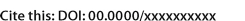} & \noindent\LARGE{\textbf{THz to far-infrared spectra of the known crystal polymorphs of Phenylalanine}} \\ 
\vspace{0.3cm} & \vspace{0.3cm} \\

 & \noindent\large{Thomas A. Niehaus,\textit{$^{a\ddag}$} Emilien Prost,\textit{$^{a\ddag}$} Vincent Loriot,\textit{$^{a}$} Franck L\'epine,\textit{$^{a}$} Luc Berg\'e,\textit{$^{b}$} and Stefan Skupin\textit{$^{a}$} }  \\

\includegraphics{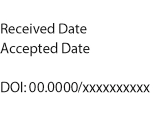} & \noindent\normalsize{There is renewed interest in the structure of the essential amino acid Phenylalanine in the solid state. Three new polymorphs were found in the years 2012 to 2014. Here, we investigate the structure, stability, and energetical ordering of these phases using first-principles simulations at the level of Density Functional Theory incorporating van-der-Waals interactions. Two of the distinct crystal forms are found to be structurally similar and energetically very close after vibrational free energy corrections have been taken into account. Infrared absorption spectra are likewise calculated and compared to experimental measurements. By combining measurements obtained with a commercial Fourier Transform Infra-Red spectrometer and a homemade air-photonics-based THz Time Domain spectrometer, we could carry out this comparison in the vibrational frequency region from 1 to 40~THz. The excellent agreement of the line positions and the established energy ranking allow us to identify the most stable polymorph of Phenylalanine.} \\

\end{tabular}

 \end{@twocolumnfalse} \vspace{0.6cm}

  ]

\renewcommand*\rmdefault{bch}\normalfont\upshape
\rmfamily
\section*{}
\vspace{-1cm}


\footnotetext{\textit{$^{a}$~Univ Lyon, Université Claude Bernard Lyon 1, CNRS, Institut Lumière Matière, F-69622 Villeurbanne, France, Tel: +33 472 431 571; E-mail: thomas.niehaus@univ-lyon1.fr}}

\footnotetext{\textit{$^{b}$~Centre Lasers Intenses et Applications, Université de Bordeaux–CNRS–CEA, 33405 Talence Cedex, France}}


\footnotetext{\ddag~These authors contributed equally to this work.}


\section{Introduction}
Over the past years, there have been significant advances in X-ray diffraction and subsequent interpretation of observed reflections in terms of the geometrical structure of the specimen \cite{fornari2009recent}. This is accompanied by an improved ability to obtain single crystals of sufficient size even for materials that are considered challenging to crystallize \cite{nagy2012advances}. One example is the essential amino acid Phenylalanine (Phe). Beginning with the measurements of Khawas \cite{khawas1970unit}, a full three-dimensional crystal structure of the D enantiomer was obtained only in 1990 by Weissbuch \cite{weissbuch1990oriented}. This was followed by the rapid discovery of different crystal structures, or polymorphs, of L-Phe. Williams found a new form free of crystal water ({\bf F-II}), starting from hydrated L-Phe in 2013 \cite{williams2013expanding}. Shortly thereafter, in 2014, Moussou reported a new monoclinic polymorph ({\bf F-III}) \cite{mossou2014self}, which was structurally similar to the Weissbuch structure but had lower symmetry. The space group assignment was already questioned earlier through Density Functional Theory calculations \cite{king2012revealing}. Also in 2014, Ihlefeldt was able to refine the Weissbuch structure, giving rise to the variant ({\bf F-I}) of the L-enantiomer and proposed a new, previously unknown polymorph ({\bf F-IV}) \cite{ihlefeldt2014polymorphs}. 

Given this rich structural diversity, it is essential to establish independent and alternative ways of structure determination and characterization. Infrared (IR) absorption spectroscopy can differentiate between phases, and provides information about subtle geometrical changes, such as modifications of hydrogen-bonding patterns. L-Phe has been well studied in the gas phase \cite{snoek2000conformational,von2008mid,ravikumar2006raman}. Still, in the solid state, only the study by Pan and co-workers using THz Time Domain Spectroscopy (THz-TDS) is available, which provided spectra for a limited frequency window from 0.5 to 4.5 THz \cite{pan2017terahertz}. 

In this work, first-principles Density Functional Theory (DFT) calculations are carried out for all known polymorphs of L-Phe, which provides previously unavailable information on the energetical ordering and stability of the various phases. We also produced simulated THz to far-IR spectra to supply vibrational fingerprints of the crystal variants. These spectra allow direct comparison to experimental data obtained using a commercial Fourier Transform Infra-Red (FTIR) spectrometer and homemade broadband THz-TDS using two-color plasma generation\cite{Kress2004,Kim2007} coupled with Air-Biased Coherent Detection (ABCD)~\cite{Karpowicz2008, Ho2010}. Using air as a generation and detection medium for THz-TDS allows extended bandwidths while avoiding the inherent phonon absorption appearing in solid-state devices.

\section{Methods}

\subsection{DFT simulations}
\label{sec-dft}
Initial crystal structures as published in \cite{ihlefeldt2014polymorphs} were retrieved from the Cambridge Structural Database (CSD) \cite{groom2016cambridge}. Polymorph {\bf F-I} at 105 K is available under the database identifier QQQAUJ05, {\bf F-I} at 298 K as QQQAUJ06, {\bf F-IV} as QQQAUJ07, and the L-Phenylalanine monohydrate {\bf F-W} as GOFWOP01. We also considered {\bf F-II} published in \cite{williams2013expanding} available as QQQAUJ04, and {\bf F-III} published in \cite{mossou2014self} and available as QQQAUJ03. The experimental cell parameters are summarized in Tab.\ \ref{tbl:cellexp}. Furthermore, the crystal structure proposed by King et al.\ \cite{king2012revealing} ({\bf F-K}) was taken from the Supplemental Information for that article. Density functional theory simulations under periodic boundary conditions were then carried out using the QUANTUM ESPRESSO code \cite{giannozzi2009quantum,giannozzi2017advanced} employing plane waves as the basis set. We used the Perdew-Burke-Ernzerhof (PBE) exchange-correlation functional together with the PAW pseudopotentials available in the library {\tt pslibrary.1.0.0}. The cutoff values for the wave functions and the density were chosen as \qty{45}{\au} and \qty{360}{\au}, respectively, and a convergence threshold of \qty{0.5e-10}{\au} was taken for the determination of the self-consistent field. Since van-der-Waals interactions are assumed to play an essential role in these partially $\pi$-stacked molecular crystals, we apply the Grimme dispersion model D3 \cite{grimme2010consistent} with default parameters. {In a benchmark of different dispersion models, this choice was recently shown to provide accurate vibrational spectra in molecular crystals also in the low THz regime\cite{bedoya2018toward}.} {The total energy computed by QUANTUM ESPRESSO contains  contributions from the kinetic energy of the electrons, the external (electron-ion), Hartree, exchange-correlation, and van-der-Waals energy, as well as the ion-ion repulsion \cite{giannozzi2009quantum,giannozzi2017advanced}. Since we are interested only in the energy differences between the different polymorphs, lattice energies are not evaluated.}  The Brillouin zone was sampled with a Monkhorst-Pack set of  2x2x1 ({\bf F-I, F-K}), 2x3x2 ({\bf F-II, F-IV, F-W}), and 2x1x2 ({\bf F-III}). The atomic positions and unit cell parameters were then optimized without any constraints using a tight convergence threshold of \qty{1.e-4}{\au}. The resulting structures maintained the initial space group in all cases. As expected, optimizing {\bf F-I} at 298~K and 105~K leads to identical structures, as the DFT simulations correspond to 0~K. The published structure for {\bf F-IV} included three different sets of atomic positions, reflecting the sidechain disorder\cite{ihlefeldt2014polymorphs}. We optimized all three structures and obtained slightly different final geometries. Only the lowest energy conformer is discussed further.
For each polymorph, the vibration modes and the THz to far-IR absorption spectrum at the $\Gamma$ point were obtained using density functional perturbation theory implemented in QUANTUM ESPRESSO. After application of the acoustic sum rule, we found no imaginary frequencies, which indicates that all final structures are indeed true minima on the potential energy surface. The spectral intensities have been Gaussian broadened with a linewidth of \qty{5}{cm^{-1}} to facilitate comparison with the experimental data. Having obtained the vibrational frequencies $\omega_i$ for mode $i$, we computed zero point energy corrected total energies $E_{ZPE} = E_{tot} + \frac{1}{2} \sum_i \hbar \omega_i $. Likewise, we determined vibrational free energy corrections  $E_{vib} = E_{tot} + F_{vib}$, with  $F_{vib} = -k_B T \sum_i\ln{Q^i_{vib}}$, $k_B$ being the Boltzmann constant, T the temperature and $Q^i_{vib}$ the partition function for mode $i$ of a quantum harmonic oscillator. {In the evaluation of the free energy corrections, the phonon dispersion should ideally be accounted for by performing a Brillouin zone integration \cite{nyman2015static}. Our results are evaluated at the $\Gamma$ point only.} {Note, that we also do not account for the thermal expansion of the crystal. As shown in Ref.~\cite{heit2016}, neglecting this effect may lead to errors up to 1~\unit{kJ/mol} in absolute terms. Since we are interested in free energy differences of polymorphs that should feature very similar expansions, we disregard it in the following. }

\begin{table}[h]
\small
  \caption{\ Experimental cell parameters of L-Phe.  }
  \label{tbl:cellexp}
  \begin{tabular*}{0.48\textwidth}{@{\extracolsep{\fill}}lllll}
    \hline
    Form & I Ref.\cite{ihlefeldt2014polymorphs} & II Ref.\cite{williams2013expanding} & III Ref.\cite{mossou2014self} & IV Ref.\cite{ihlefeldt2014polymorphs}\\
    \hline
    space group               & P2$_1$  &   P2$_1$  & P2$_1$     &    C2  \\
    a [\AA]                   &  8.7829(4) &    12.063(11) &     6.0010(5) &     9.6806(7) \\
    b [\AA]                   &  5.9985(3) &     5.412(5) &     30.8020(17) &     5.2362(4) \\
    c [\AA]                   & 31.0175(13) &    13.676(13) &     8.7980(4) &    15.8474(12) \\
    $\beta$ [$\mbox{}^\circ$] & 96.9220(12) &    99.5976(26) &    90.120(4) &    96.291(3) \\
    V  [\AA$^3$]              & 1622.12(13) &   880.3(24)    &  1626.24(17)&   798.46(10)\\
    Z'      & 4 & 2 & 4 & 1 \\
    {T [K]}                     & {105}     &   {294}      &  {100}      &   {100}  \\
    \hline
  \end{tabular*}
\end{table}

\subsection{Experimental setups}
The FTIR measurements were performed using a commercial Bruker Invenio-S spectrometer. This machine uses a polychromatic light source that works in the range of 400~\unit{cm^{-1}} to 4000~\unit{cm^{-1}}. The light goes through a Michelson-type interferometer before being directed through the sample and finally detected. Data acquisition is performed by scanning the distance between the two arms of the interferometer and recording the resulting interferogram. For each position of the interferometer, interference leads to a change in the spectral profile of the beam. It is possible to extract this information by taking the Fourier transform of the interferogram. Given that this interferogram measures the light intensity versus a position corresponding to a time delay between the two arms, we directly obtain the spectral profile in reciprocal space. 

To extend the experimentally accessible frequency range below 400~\unit{cm^{-1}}, we use a homemade air-photonics-based THz Time Domain Spectroscopy (THz-TDS) system relying on two-color plasma THz generation. Ultrashort optical pulses of 30~fs at 800~nm produced by a Ti:Sa amplifier are focused in the laboratory atmosphere. A BBO crystal, placed between the focusing lens and the geometrical focus, produces a second color by type I second harmonic generation. An ultra-thin half-waveplate at 800 nm placed after the BBO crystal is used to align the polarization of both colors. Near the geometrical focus, an air plasma is generated by the two-color laser pulse. This plasma emits THz radiation, collected and steered by off-axis parabolic mirrors. The THz field is first focused on the sample and then refocused on the THz Air-Biased Coherent Detection (THz-ABCD) detection system. This detection scheme overlaps the THz field with an IR probe pulse and a high-voltage bias field. Then, second harmonic (SH) radiation is generated through four-wave mixing in air. The SH signal produced is proportional to the THz and bias electric field, and by taking successive measurements for opposite signs of the bias field, one can extract the THz electric field. The THz field temporal profile is then reconstructed by scanning the delay between the THz pulse and the IR probe pulse. Fourier transform gives access to the THz spectrum. Our setup allows us to perform THz-TDS measurements between 4~THz and 30~THz. More details on the specific implementation of our THz-TDS system can be found in \cite{Prost2022RSI}.

\subsection{Sample preparation and measurements}\label{sec:sample_prep}
High-purity D-Phenylalanine powder ($\ge$98$\%$, Sigma-Aldrich) was used in our experiments. The powder grain size is of the order of the wavelengths ($\sim~10-300~\mathrm{\mu}$m) in our frequency window, thus diffusing the THz radiation. Therefore, sample preparation was required for the powder to be used with our apparatus. 
{We chose to deposit a thin film of Phe on a silicon wafer substrate by physical vapor deposition. The Phe powder was heated to 260°C to evaporate under vacuum while the substrate was kept at room temperature, allowing the molecules to condensate on the substrate. We obtained Phe films of about 10~µm in thickness.}

The silicon wafer used as the substrate induces a Fabry-Pérot effect because of multiple Fresnel reflections at its interfaces.
This has been filtered out with THz-TDS by restricting the time domain to delays below the first back-and-forth reflection in the substrate. For FTIR, however, it led to oscillations in the spectra that were removed by Fourier filtering. This filtering led to a floating baseline in the absoption signal. Therefore, the discussion is focused on the positions of the absorption and relative amplitudes unaffected by these artifacts. Finally, reference measurements employing a silicon wafer without deposition were performed to extract the pure D-Phe transmission spectra from our THz-TDS and FTIR measurements. 

\section{Results and discussion}
\subsection{Calculated structure and energetics}
\begin{figure*}[htb]
\centering
  \includegraphics[width=\textwidth]{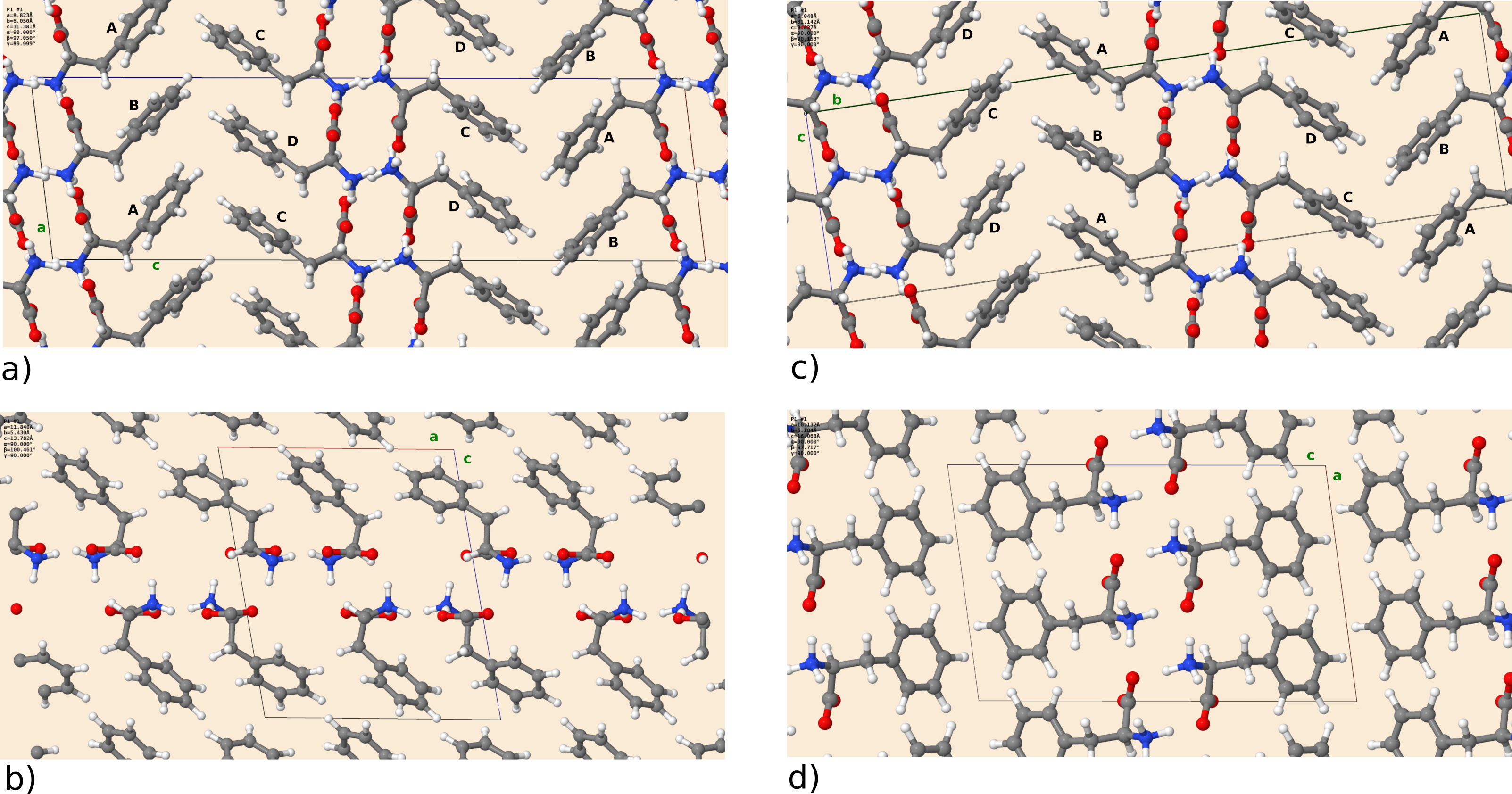}
  \caption{DFT optimized crystal structures of L-Phe for polymorph a) {\bf F-I}, b) {\bf F-II}, c) {\bf F-III}, and d) {\bf F-IV}. The unit cell is depicted as a box, and the crystallographic directions (a,b,c) are marked by green letters. {\bf F-I} and {\bf F-III} have four distinct L-Phe units, which are labeled A to D.}
  \label{poly}
\end{figure*}

First, we will briefly discuss the relaxed crystal structures shown in Fig.~\ref{poly}. All polymorphs are characterized by hydrogen-bonded bilayers and hydrophobic regions given by the phenyl sidechains.  {\bf F-I} and {\bf F-III} are structurally highly similar and contain four molecules $A$ to $D$  in the asymmetric unit. Monomers differ in the dihedral angle of the phenyl rings ({\bf F-I}: A 93.5$^\circ$, B 59.7$^\circ$, C 57.3$^\circ$, D 96.0$^\circ$) and ({\bf F-III}: A 94.0$^\circ$, B 59.7$^\circ$, C 57.4$^\circ$, D 95.8$^\circ$), with $A$ being similar to $D$ and $B$ similar to $C$. As visible in Fig.~\ref{poly}, the two crystal forms can only be differentiated by their monomer arrangement in the hydrophobic region. {Transforming {\bf F-I} into {\bf F-III} would require a shift of every other bilayer along the $a$ axis. Cuppen and co-workers investigated this phase transition in detail \cite{cuppen2019rich}, but found no experimental evidence of its occurence in the temperature range from 100 K to ambient temperature. Interestingly, though, they reported that polymorph {\bf F-I} could no longer be crystallized at room temperature after growth of {\bf F-III}. This indicates that {\bf F-III} is the kinetically preferred form. Further,} Polymorph {\bf F-II} was found by Williams \cite{williams2013expanding} under dry conditions in a study of L-Phe hydrates. The open structure with a smaller density than {\bf F-I}/{\bf F-III} can host additional water molecules. The recent polymorph {\bf F-IV} is the most compact and features a different stacking with respect to {\bf F-I}/{\bf F-III} \cite{ihlefeldt2014polymorphs}. The DFT calculations can reproduce the experimental lattice parameters satisfactorily. Table~\ref{cellexp} shows that the deviation does not exceed 2\% for {\bf F-I} to {\bf F-III}, with larger differences for {\bf F-IV}. As reported in Ref.~\cite{ihlefeldt2014polymorphs}, the latter structure exhibits a large side chain disorder, making structural refinement more difficult. Given that for the studied molecular crystals rather small changes in sidechain stacking can lead to substantial changes in the lattice parameters, van-der-Waals interactions seem well described by the Grimme dispersion model employed here. The optimized crystal structures have been deposited on the Zenodo repository \cite{zenodo}. 

Turning now to the energetical ordering of the different phases indicated in Tab.~\ref{ene}, one observes that the high structural similarity of {\bf F-I} and {\bf F-III} translates into a very small energy difference of $\Delta E_{tot}$ =  0.186~\unit{kJ/mol}. Taking into account the zero point energy (ZPE) and vibrational entropy corrections, the difference even reduces to  $\Delta E_{vib}$  = 0.064~\unit{kJ/mol} at $T = 300$ K. This indicates that both phases should be present in L-Phe samples in equal amounts unless special care is taken during the crystallization. {Our results are in line with the work of Cuppen et al.\cite{cuppen2019rich} that also suggests a higher stability of {\bf F-III} than {\bf F-I}. Given the very small energy difference and the inevitable approximations in our DFT approach, a final verdict on the thermodynamic stability of both phases cannot be given at this point.} We briefly comment on phase {\bf F-K} found by King et al.\ \cite{king2012revealing} via DFT simulations. Although these authors refer to {\bf F-K} as the "correct crystal structure of L-phenylalanine," {we find it to be higher in energy ($\Delta E_{tot}$ =  0.296~\unit{kJ/mol}) and so far it was not detected in crystallography. Therefore, we do not discuss it further.} The experimentally confirmed phases {\bf F-II} and {\bf F-IV} are even higher in energy. In the case of {\bf F-II}, this could be due to the more porous structure leading to reduced dispersion interactions between the phenyl rings. The high energy of {\bf F-IV} is surprising, as this conformer was speculated to be the lowest energy phase in Ref.~\cite{ihlefeldt2014polymorphs} because of its high density. {So far, however, the large structural disorder of this phase was not taken into account (see section \ref{sec-dft}). The three experimentally available structures give rise to configurational entropy ($S_{conf} = -k_B \sum_i p_i \ln(p_i)$ with $p_i\propto \exp(-E_{tot,i}/k_B T)$), which we can estimate from the DFT energies for the respective structures. The corresponding free energy contribution of -2.273~\unit{kJ/mol} at $T = 300$ K is sizable and stabilizes {\bf F-IV}, but does not change the general energy ranking. Furthermore, the experimentally derived occupancies for the three structures (0.471:0.260:0.267) do not match well with the probabilities $p_i$ obtained from the DFT calculations (0.501:0.079:0.420). A larger deviation of theory and experiment for phase {\bf F-IV} was already mentioned in the context of the lattice parameters and deserves further study.}

\begin{table}[htb]
\small
  \caption{\ Percent deviation [$100 (x_{cal} - x_{exp})/x_{exp}$] of calculated lattice parameters from the experimental ones listed in Tab.\ \ref{tbl:cellexp}}
  \label{cellexp}
  \begin{tabular*}{0.48\textwidth}{@{\extracolsep{\fill}}lllll}
    \hline
    Form & I & II & III & IV \\
    \hline
    a   &      0.455 &     -1.848 &      0.791 &      4.658            \\
    b   & 0.861 &      0.341 &      1.105 &     -1.000               \\
    c   &   1.173 &      0.777 &      0.333 &      1.390              \\
    $\beta$  &  0.132 &      0.867 &      0.036 &      1.481\\
    V    &  2.486 &     -1.008 &      2.245 &      4.730               \\  
    \hline
  \end{tabular*}
\end{table}
\begin{table}[h]
\small
  \caption{\ Relative energies in \unit{kJ/mol} with respect to the lowest energy structure. $\Delta F_{ZPE}$ contains the vibrational zero point energy and $\Delta E_{vib}$  contains entropy corrections at $T = 300$ K.}  
  \label{ene}
  \begin{tabular*}{0.48\textwidth}{@{\extracolsep{\fill}}llll}
    \hline
     & $\Delta E_{tot}$ & $\Delta E_{ZPE}$ & $\Delta E_{vib}$  \\
    \hline
    {\bf F-I}   & 0.186   &   0.144   &   0.064      \\
    {\bf F-II}  &   4.452  &   4.442  &   6.796 \\
     {\bf F-III} &  0.000   &  0.000   &  0.000 \\
      {\bf F-IV} &  6.400   &  6.542   &  7.706 \\
      {\bf F-K} &  0.296 & N/A & N/A\\
    \hline
  \end{tabular*}
\end{table}

\subsection{THz to far-IR spectra}
\begin{figure*}[htb]
\centering
  \includegraphics[width=0.45\textwidth]{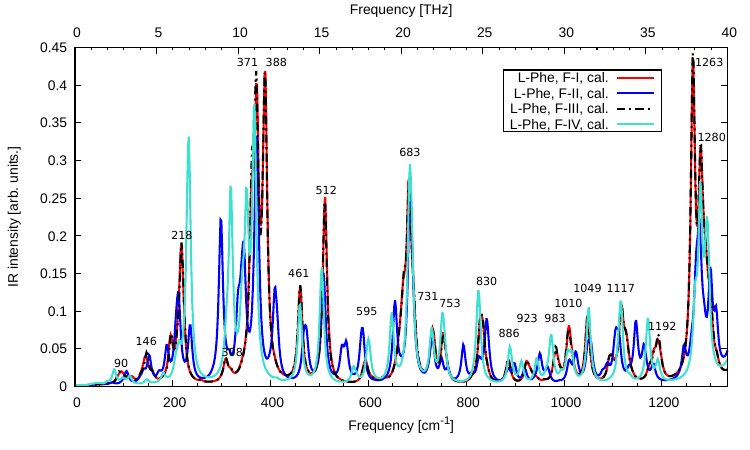}
   \includegraphics[width=0.45\textwidth]{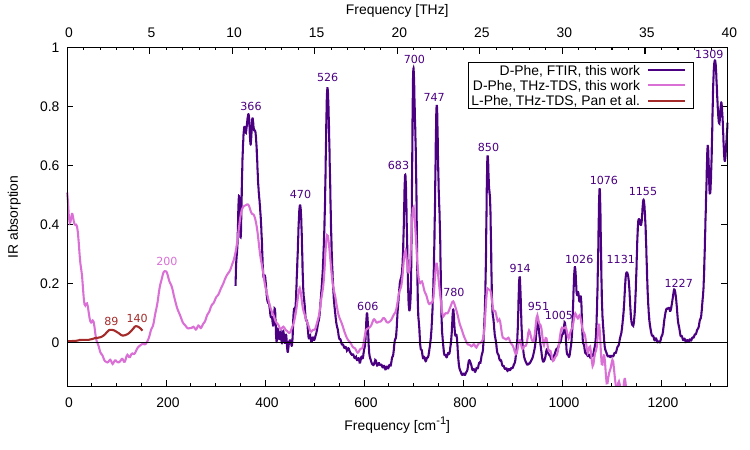}
  \caption{(Left, theory) Calculated infrared absorption spectrum for L-Phe polymorphs {\bf F-I} to {\bf F-IV}. Spectra are normalized to account for the different numbers of monomers in the unit cell. (Right, experiment) FTIR and THz-TDS infrared absorption spectrum for D-Phe together with the results for L-Phe from Pan et al. \cite{pan2017terahertz}. }
  \label{spec}
\end{figure*}
We compare the simulated THz to far-IR absorption spectra of {\bf F-I} to {\bf F-IV} with the THz-TDS results from Pan \textit{et al.}\ \cite{pan2017terahertz} and our measurements in Fig.~\ref{spec}. Although our experimental sample consisted of the enantiomer D-Phe, we compare it here to calculations for L-Phe. In the framework of this study, both enantiomers are equivalent because conventional infrared spectroscopy involving linearly polarized radiation is not sensitive to chirality. The first observation that can be made is that {\bf F-I} and {\bf F-III} are not only structurally and energetically close, but their vibrational spectrum is also nearly identical. This holds not only in the depicted region up to 40~THz but also for higher frequencies (see Fig.~\ref{800bis}). The general agreement between the experimental results and the simulations is highest for {\bf F-I}/{\bf F-III}, which confirms that they are indeed the most stable polymorphs and the main constituents of the sample. In Fig.\ref{spec}, the low energy region of the spectrum measured by our THz-TDS spectroscopy setup is not well described because of incomplete spatial overlap between the THz field and the probe pulse in the THz-ABCD. This suppresses the THz signal at the lowest frequencies\cite{Prost2022EPJST} up to approximately 150~\unit{cm^{-1}}. 
Therefore, we compare to the experimental results of Pan et al.\cite{pan2017terahertz} in this frequency range. There are two main absorption peaks at 89~\unit{cm^{-1}} and 140~\unit{cm^{-1}}. Using the eigenvectors of the dynamical matrix for the {\bf F-III} simulations, we can assign these modes to wagging (90~\unit{cm^{-1}}) and twisting (146~\unit{cm^{-1}}) of phenyl rings, respectively. We further identify the {\bf F-III} peak at 218~\unit{cm^{-1}} as a hydrogen bond stretching motion $\nu(\text{O-NH}_3)$. Experimentally, this peak appears around 200~\unit{cm^{-1}}. In the range 250-400~\unit{cm^{-1}}, {\bf F-III} features three important rocking modes, at 308~\unit{cm^{-1}} ($\rho(\text{CH}_2)$), 371~\unit{cm^{-1}} ($\rho(\text{NH}_3)$) and 388~\unit{cm^{-1}} ($\rho(NH_3)$). The intensity of the latter two modes is overestimated in the simulations compared to the FTIR measurements,  where a broad feature is found at 366~\unit{cm^{-1}}. The forms {\bf F-II} and {\bf F-IV} show large deviations from the {\bf F-I}/{\bf F-III} spectrum, which is due to the different hydrogen-bonded network of the hydrophilic bilayer in these structures. Further important peaks in the range up to 800~\unit{cm^{-1}} are found at 461~\unit{cm^{-1}}, 512~\unit{cm^{-1}} and 683~\unit{cm^{-1}}, all due to out-of-plane motions of the phenyl rings, which fairly match the corresponding experimental absorption features at 470~\unit{cm^{-1}}, 526~\unit{cm^{-1}} and 700~\unit{cm^{-1}}. The intensity of the mode at 683~\unit{cm^{-1}} seems to be overestimated in the simulations. We attribute this fact to the underestimated dispersion of this mode. In fact, the four molecules in the unit cell lead to nearly identical normal modes (682~\unit{cm^{-1}} to 685~\unit{cm^{-1}}) with a strong absorption intensity, giving rise to one large peak. We speculate that the FTIR excitations at 683 and 700 ~\unit{cm^{-1}} derive from the same mode, indicating a much larger dispersion and leading to well separated peaks with smaller absorption.

Above this frequency range, notable absorption can be found at 850~\unit{cm^{-1}} (wagging of the phenyl hydrogens, calculation at 830~\unit{cm^{-1}}), 1076~\unit{cm^{-1}} (phenyl stretching, calculation at 1049~\unit{cm^{-1}}), 1155~\unit{cm^{-1}} (wagging of the amino groups, calculation at 1117~\unit{cm^{-1}}) and strong absorption around 1309~\unit{cm^{-1}} (C-O stretch, calculation at 1263~\unit{cm^{-1}}). A detailed list of all calculated and measured modes is given in Tab.~\ref{modes}. We find that FTIR and THz-TDS spectroscopy provide identical peak positions (except for the mode at 366~\unit{cm^{-1}}), while the spectral intensity is more reliably given at the FTIR level. This holds in particular above 900~\unit{cm^{-1}} where the bandwidth limit of the THz-TDS is reached. The mean absolute error of the calculated frequencies for {\bf F-III}  with respect to the FTIR data is 24~\unit{cm^{-1}}. Scaling the frequencies by a factor of 1.03 to account for systematic errors in the DFT calculation reduces the error to only 6~\unit{cm^{-1}}. Such scaling is routinely performed for simulations in the gas phase \cite{merrick2007evaluation} but is less common for molecular crystals, though it works quite well in the present example.

Simulations of the monohydrate {\bf F-W} (see Fig.~\ref{bis800} and \ref{800bis}) provide a complex spectrum significantly different from the experimental results and simulations. This is also true for {\bf F-II}, structurally similar to {\bf F-W}, but without the additional water molecule. This indicates that the studied samples contain pure D-Phe, as expected.      

\section{Conclusions}
Extensive DFT simulations of all known crystal polymorphs of Phenylalanine have been carried out. Experimentally known lattice parameters are well reproduced, and van-der-Waals interaction seems well described by the Grimme dispersion model. An energetical ordering of the different phases is established, where the polymorphs ({\bf F-I}), Ref.\cite{ihlefeldt2014polymorphs}, and ({\bf F-III}), Ref.\cite{mossou2014self}, feature the lowest energies. They are found to be very close structurally and energetically, and their simulated vibrational spectra are nearly identical. They best agree with our experimental spectra obtained with the FTIR and THz-TDS methods in the frequency range of 1 to 40~THz from our thin-film deposition sample. This confirms that ({\bf F-I}) and ({\bf F-III}) are indeed the most stable polymorphs of Phenylalanine. Our work demonstrates how advanced numerical and experimental techniques can be used in concert to reveal the structure of complex organic molecules in the solid state. We believe that detailed structural information on the amino acid Phenylalanine obtained in our study will be helpful for further research.

\section*{Acknowledgements} \label{sec:acknowledgements}
We acknowledge the Agence Innovation Defense (AID) and Agence Nationale de la Recherche (ANR-19-ASMA-0007-ALTESSE2) for providing financial support and thank GENCI for computational resources under projects DARI A0110810637 and A0130810637.
We thank Eric Constant and Isabelle Compagnon for stimulating discussions.


\balance


\bibliography{Combined} 
\bibliographystyle{rsc} 

\appendix

\begin{figure*}[h]
\centering
  \includegraphics[width=0.8\textwidth]{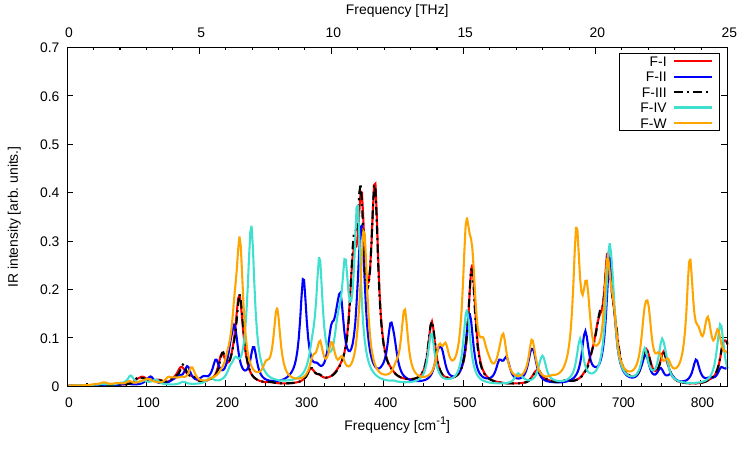}
  \caption{Calculated absorption spectrum of all considered polymorphs {\bf F-I} to  {\bf F-IV} together with the monohydrate   {\bf F-W}  up to 25 \unit{THz}. The intensity is normalized for the number of molecules in the simulation cell.} 
  \label{bis800}
\end{figure*}

\begin{figure*}[h]
\centering
  \includegraphics[width=0.8\textwidth]{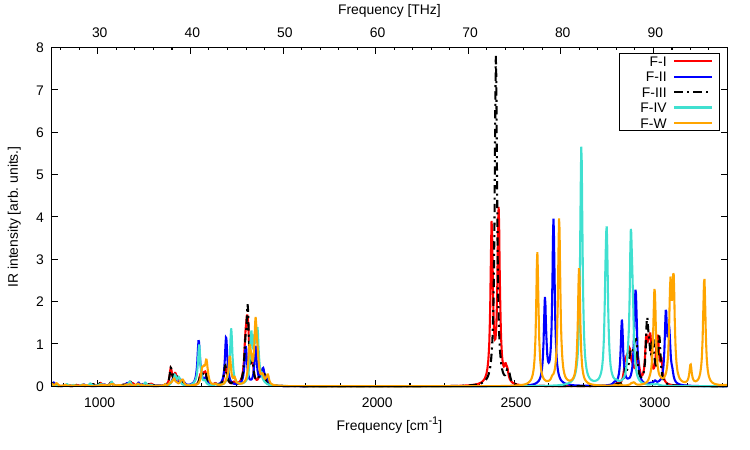}
  \caption{Calculated absorption spectrum of all considered polymorphs {\bf F-I} to  {\bf F-IV} together with the monohydrate   {\bf F-W}  between 25 and 100 \unit{THz}. The intensity is normalized for the number of molecules in the simulation cell.} 
  \label{800bis}
\end{figure*}

\begin{table*}
    \centering
    \begin{tabular}{cccc}
     \hline
    theory ({\bf F-III}) &  experiment (THz-TDS) & experiment (FTIR) & mode description  \\\hline
    90 & & &$\omega$(Phe)\\
    146 & & &$\tau$(Phe)\\
    158 & & &$\rho$(Phe)\\
    197 & & & $\rho$(Phe)\\
    218 & 200 & & $\nu(\text{O-NH}_3)$\\
    308 &  & & $\rho(\text{CH}_2)$\\
    371 & 370 & 366& $\rho(\text{NH}_3)$\\
    388 & & 374&$\rho(\text{NH}_3)$ \\
    461 & 470 & 470& $\gamma$(Ph)\\
    512 & 526 & 526& $\gamma$(Ph)\\
    595 &606 & 606& $\tau$(NH$_3$)\\
    683 &700 & 700&$\gamma$(Ph)\\
    731 &747 & 747& $\omega$(H-Ph)\\
    753 & 780& 780& $\gamma$(CO$_2$)\\
    830 & 853 & 850 & $\omega$(H-Ph)\\
    886 & 914 & 914& $\tau$(H-Ph)\\
    923 & & 951& $\delta$(CH-NH$_3$)\\
    983 & & 1005 & $\omega(\text{NH}_3)$\\
    1010 & & 1026& $\nu$(Ph)\\
    1049 & & 1076 & $\nu$(Ph)\\
    1096 & & 1131& $\nu$(C-N)\\
    1117 & & 1155& $\omega(\text{NH}_3)$\\
    1192 & & 1227& $\nu$(C-C)\\
    1263 & & 1309& $\nu$(C-O)\\
    1280 & & 1322& $\nu$(C-C)\\
    \end{tabular}
    \caption{Vibrational frequencies of Phenylalanine up to 1300 cm$^{-1}$ from the DFT calculations for conformer {\bf F-III} and the two experimental techniques (THz-TDS, FTIR) used in this article in units of wavenumbers (cm$^{-1}$). The abbreviations in the approximate mode description stand for: $\tau$: twisting, $\rho$: rocking, $\omega$: wagging, $\nu$: stretching, $\delta$: scissoring, $\gamma$: out-of-plane motion, Ph: phenyl ring, Phe: Phenylalanine.}
    \label{modes}
\end{table*}

\end{document}